\documentclass[prl,aps,twocolumn,amsmath,amssymb]{revtex4}
\usepackage{graphicx}
\usepackage{color}
\usepackage{ulem}

\newcommand{\be}{\begin{equation}}
\newcommand{\ee}{\end{equation}}

\newcommand{\bex}{\begin{eqnarray}}
\newcommand{\eex}{\end{eqnarray}}
\newcommand{\bmin}{\begin{center}\begin{minipage}{460pt}}
\newcommand{\emin}{\end{minipage}\end{center}}

\begin{document}

\title{A  general  criterion  for  nonclassicality  from  a  signaling
  perspective}

\author{S. Aravinda} 

\affiliation{Poornaprajna    Institute    of   Scientific    Research,
  Sadashivnagar, Bangalore, India}

\author{R.  Srikanth}
\email{srik@poornaprajna.org}  

\affiliation{Poornaprajna    Institute    of   Scientific    Research,
  Sadashivnagar, Bangalore, India}
\affiliation{Raman   Research  Institute,   Sadashivnagar,  Bangalore,
  India.}

\begin{abstract}
We  argue   that  the  essence  of  nonclassicality   of  a  bipartite
correlation  is a  positive  signal deficit--  the communication  cost
excess  over  the  available  signaling.   By  this  criterion,  while
violations of Bell-type and contextuality inequalities are necessarily
non-classical,  some  violations of  the  Leggett-Garg inequality  are
classical.    Further,  signaling   tends  to   diminish  nonclassical
properties, such as intrinsic randomness, no-cloning, complementarity,
etc.  Signal deficit is shown to have its ultimate origin in intrinsic
randomness.    A   possible   analogy   of  nonclassicality   to   the
metamathematical concept of G\"odel incompleteness is noted.
\end{abstract}

\maketitle

\paragraph{Introduction.}
What exactly makes  quantum mechanics (QM) nonclassical? Traditionally
this question has  been answered in different ways  in quantum optics,
in  the foundations  of QM,  etc.  In  quantum information  theory, we
associate    nonclassicality    with    features   like    fundamental
indeterminism,  Heisenberg   uncertainty,  monogamy  and   privacy  of
correlations, and  the impossibility of perfect cloning.   In terms of
bi-partite correlations,  it turns out that all  these features derive
from just  two assumptions  \cite{mag06}.  The first  is no-signaling,
which stipulates that signals cannot propagate except through material
mediation.    The   second   is  nonlocality,   whereby   multipartite
correlations can  violate Bell-type inequalities \cite{bell64,chsh69}.
It is  known that a  wider class of generalized  probability theories,
e.g.,  the  Popescu-Rohrlich   (PR)  box  \cite{pr94},  satisfy  these
postulates.

Yet, nonclassicality is  arguably strictly weaker.  Local correlations
in QM,  such as that between  sequential measurements on  a qubit, can
violate the  temporal equivalent  of the Bell-type  inequality, namely
the   Leggett-Garg  (LG)  inequality   \cite{lg85},  but   such  local
correlations involve `signaling in time', so that the above postulates
do not cover such phenomena.   We consider in this article the problem
of refining  the above criterion  of non-classicality by  relaxing the
no-signaling  condition.   (This  does  not violate  the  relativistic
prohibition   on  superluminal   signaling,   because  the   signaling
correlations  considered here  are between  timelike  separated events
occuring on the same particle, and thus just consistute memory.)

Since   classical  models   that   simulate  nonlocal,   non-signaling
correlations entail  non-vanishing communication cost  \cite{tb03}, it
is natural  to suppose that  the weakening of  nonclassicality through
signaling can be characterized in terms a signaling deficit: i.e., the
excess   of  communication  cost   for  a   given  \textit{correlation
  inequality} (Bell, contextuality or  LG) above the available signal.
The remaining article is devoted to the elucidation of this point

\paragraph{Bell and LG inequalities.} 
The  LG inequality  is satisfied  by all  noninvasive-realist theories
(For  related variants,  cf.  Ref.   \cite{lap06}).  `Realism'  is the
assumption that the given  system $Q$ possesses determinate properties
prior  to   measurement.   `Noninvasiveness'  is   the  assumption  of
measurability of a system  without disturbing the subsequent evolution
of its possessed value.  Thus, measurement only reveals a pre-existing
value.

Suppose the quantities labeled $a=0, 1$ are measured at time $t_A$ and
those labeled  $b=0,1$ at time  $t_B > t_A$, both  yielding respective
outcomes  $x,y =  \pm  1$.   Then under  the  stated assumptions,  any
correlation $\textbf{P} \equiv P(x,y|a,b)$ satisfy:
\begin{equation}
\Lambda(\textbf{P}) \equiv
|\sum_{a,b} (-1)^{a\overline{b}}(xy)| \leq 2,
\label{eq:lg}
\end{equation}
which  is  the  LG  inequality  in its  2-time  ($t_A,  t_B$)  variant
\cite{bruk04}. At  the microscopic scale,  because quantum measurement
is invasive, the LG inequality was originally proposed for macroscopic
systems, which can in  principle be measured non-invasively.  When $a$
and  $b$  are  interpreted   as  observables  belonging  to  spatially
separated      particles,      Eq.       (\ref{eq:lg})     is      the
Clauser-Horne-Shimony-Holt   (CHSH)    inequality   \cite{chsh69},   a
Bell-type inequality.

The assumptions  behind the derivation  of a Bell-type  inequality are
localism and realism.  As a  classical theory is necessarily local and
realist,  a violation of  Bell's inequality  implies non-classicality.
Likewise,  as a  classical  theory is  necessarily non-contextual  and
realist, a  violation of a contextuality  inequality \cite{bad09} also
implies non-classicality.  However, the violation of the LG inequality
does  not entail  non-classicality, since  a classical  theory  is not
necessarily non-invasive.

Invasiveness implies  a signal \cite{fritz10} carried  forward in time
from one  measurement to another on  the same particle,  such that the
probability distribution  of a  subsequent measurement depends  on the
choice made  earlier \cite{kofbruk12}. For  sufficiently large signal,
an  invasive-realist classical  mechanism  can presumably  be used  to
violate the LG  inequality.  Then, the violation of  the LG inequality
is not  non-classical unless  the degree of  violation is shown  to be
larger than  can be explained by  a classical mechanism  that uses the
signaling in the correlations.

A  set  of correlations  $\textbf{P}  \equiv  P(x,y|a,b)$ between  two
parties $\mathcal{A}$  and $\mathcal{B}$ can  be described by  a joint
distribution if and  only  if  it  has a  deterministic  hidden
variable    (HV)   description    $P(x,y|a,b)    =   \int\rho(\lambda)
P(x|a,\lambda) P(y|b,\lambda) d\lambda  $ \cite{fine82}.  For the case
where  $a, b  =  0,1$  and $x,  y  = \pm  1$,  this  is equivalent  to
satisfying correlation inequality  (\ref{eq:lg}) (no matter whether it
is spatial, temporal or context-based).

Violation  of a  correlation  inequality constitutes  what  we call  a
general  \textit{disturbance}.  Nonlocality  or  contextuality is  one
type   of    disturbance.    Disturbance    may   or   may    not   be
\textit{signaling}.   In QM,  if the  joint measurements  $a$  and $b$
commute  (as  in  Bell-type  or contextuality  inequality),  then  the
disturbance is non-signaling.

\paragraph{Signaling.}

Let $P_0^{b}$ ($P_1^{b}$, resp.)   represent the probability that Bob,
measuring observable $b$, finds  $y=+1$ when Alice measures observable
$a=0$ ($a=1$, resp.).  A  qualitative indication of signal strength is
$s   \equiv  \max_{b}|P^b_0   -  P^b_1|   >  0$,   where   $b$.   More
quantitatively, the signalled  information $S(\textbf{P})$ received by
Bob is  quantified by the  mutual information $I(A:Y)$  maximized over
Alice's and  Bob's choices.  Letting  $\alpha$ ($\beta\equiv1-\alpha$)
denote the probability with  which Alice chooses $a=0$ ($a=1$), $Q_j^b
\equiv 1  - P_j^b$, $\overline{P}  \equiv \alpha P_0^b +  \beta P_1^b$
and  $\overline{Q}   \equiv  \alpha  Q_0^b  +  \beta   Q_1^b$,  it  is
straightforward to find (cf. Appendix)
\begin{widetext}
\begin{eqnarray}
S(\textbf{P}) =  \max_{b,\alpha} [ H(\alpha) +  \alpha P_0^b \log
  ( \alpha P_0^b/\overline{P}) + \beta P_1^b  \log(
  \beta P_1^b/\overline{P}) + \alpha Q_0^b \log  (\alpha
  Q_0^b/\overline{Q}) +  \beta  Q_1^b  \log(\beta
  Q_1^b/\overline{Q})].
\label{eq:chiX}
\end{eqnarray}
\end{widetext}

\paragraph{Communication cost.}
Disturbance implies that the communication cost, the information about
$a$ required to output $y$ (or about $b$ to output $x$) for simulating
$\textbf{P}$,  is greater  than  0.  Following  Ref. \cite{pir03},  we
consider the general \textbf{P}  as a mixture of \textit{deterministic
  strategies}.   Examples of such  strategies are  deterministic local
correlations   like  $\textbf{d}^{0_0}$   defined  by   $P(x,y|a,b)  =
\delta^x_0\delta^y_0$        and        $\textbf{d}^{4_0}       \equiv
\delta^x_a\delta^y_b$.   For  these the  signal  (\textbf{S}) and  the
communication  cost (\textbf{C})  are identically  zero.  Correlations
for   which  $S(\textbf{P})  =   C(\textbf{P})  =   1$  bit   are  the
deterministic   1-bit   strategies   $\textbf{d}^{0_1}$   defined   by
$P(x,y|a,b)   =  \delta^x_0\delta^y_{a\cdot  \overline{b}}$   and  its
\textit{signal   complement}   $\textbf{d}^{3_1}   \equiv   \delta^x_1
\delta^y_{a\cdot\overline{b}+1}$:  a convex  combination of  these two
reduces the  signal at  a fixed  communication cost of  1 bit,  and at
fixed maximal inequality violation of $\Lambda(\textbf{P}) = 4$ in Eq.
(\ref{eq:lg}).   Indeed,  $S(\textbf{d}^{0_1})  =  1$ bit,  being  the
maximal signal from Alice to Bob (when she chooses $a=0$ or $a=1$ with
equal   probability   and  Bob   always   chooses  $b=1$).    Further,
$C(\textbf{d}^{0_1}) = 1$ since Bob needs $\log|a|$ bits (here: 1 bit)
of   communication  specifying   $a$,  in   order  to   output   $y  =
a\cdot\overline{b}$.

For Eq.   (\ref{eq:lg}), it can be  shown (cf. the  Appendix) that the
average communication cost
\begin{equation}
C(\textbf{P}) = \max\left(C_\Lambda(\textbf{P}),S(\textbf{P})\right),
\label{eq:piro}
\end{equation}
where  $C_\Lambda(\textbf{P}) \equiv  \frac{1}{2}\Lambda(\textbf{P}) -
1$ is referred to  as the \textit{disturbance cost}.  This generalizes
the result of Ref.  \cite{pir03}, where $C_\Lambda(\textbf{P})$ is the
general  lower  bound  on  average communication  cost.   For  maximal
quantum   violation  of   the  inequality   (\ref{eq:lg}),   which  is
$2\sqrt{2}$, $C_\Lambda(\textbf{P}) = \sqrt{2}-1 \approx 0.41$ bits.

For a probabilistic \textbf{P} obtained  by mixing the local and 1-bit
deterministic strategies, by virtue of positivity of $C$ and convexity
of $S$, we have:
\begin{equation}
S(\textbf{P}) \le  C(\textbf{P}),
\label{eq:metamath}
\end{equation}
This is  also generally true,  since $S(\textbf{P}) \equiv  I(a:y) \le
H(a) \equiv  C(\textbf{P})$, where $I$  is mutual information  and $H$
binary    entropy.    A   uniform    mixture   consisting    only   of
$\textbf{d}^{0_1}$   and   $\textbf{d}^{3_1}$   yields  the   PR   box
\cite{pr94}, for which $C(\textbf{P})=1$, but $S(\textbf{P})=0$.

\textit{Quantum  correlations.}  In  what follows,  we may  denote the
observables    $a=0,1$     ($b=0,1$)    by    $\hat{a},\hat{a}^\prime$
($\hat{b},\hat{b}^\prime$).  For  a qubit in QM,  the correlations for
sequential  measurements   $\hat{a}$  then  $\hat{b}$   are  given  by
$P(x,y|\hat{a},\hat{b})  =   \textrm{Tr}\left(\frac{1  +  y\hat{b}}{2}
\frac{1 + x\hat{a}}{2}\rho \frac{1 + x\hat{a}}{2}\right) = \frac{1}{4}
+                 \frac{x}{4}\textrm{Tr}(\hat{a}\rho)                +
\frac{y}{8}\textrm{Tr}(\hat{b}\rho)                                   +
\frac{xy}{8}\textrm{Tr}(\{\hat{a},\hat{b}\}\rho)                      +
\frac{y}{8}\textrm{Tr}(\hat{a}\hat{b}\hat{a}\rho)$.   Given that Alice
measures $\hat{a}$ or $\hat{a}^\prime$, the condition for signaling is
$s  \equiv  |P_{\hat{a}}^{\hat{b}}  - P_{\hat{a}^\prime}^{\hat{b}}|  =
\left|  \frac{1}{4} \textrm{Tr}  \left[  \left(\hat{a}\hat{b}\hat{a} -
  \hat{a}^\prime\hat{b}\hat{a}^\prime\right)\rho \right] \right| > 0$,
which   is   bounded  above   in   QM   as:   $  s   \le   \frac{1}{4}
\left|\textrm{Tr}\left(\hat{a}\hat{b}\hat{a}\rho\right)\right|        +
\frac{1}{4}
\left|\textrm{Tr}\left(\hat{a}^\prime\hat{b}\hat{a}^\prime\rho\right)
\right|  \le  \frac{1}{2}$.  Thus,  $[\hat{a},\hat{b}]  \ne 0$  and/or
$[\hat{a}^\prime,\hat{b}] \ne 0$ is necessary for signaling.

Let   the   density   operator   obtained   by   measuring   $\hat{a}$
($\hat{a}^\prime$)   on   $|\psi\rangle$   be   $\rho_0$   ($\rho_1$).
Intuitively, we  expect to maximize signaling  $s$ when $|\psi\rangle$
is an eigenstate of $\hat{a}$, and $\hat{a}^\prime$ maximally fails to
commute  with $\hat{a}$  (i.e., the  two  observables form  a pair  of
\textit{mutually unbiased  bases}).  By a similar argument  as used to
bound   $s$,   trace    distance   $\tau$   satisfies   $\tau   \equiv
\frac{1}{2}||\rho_0 - \rho_1||  \le \frac{1}{2}$.  This inequality, as
well as  that for $s$ above,  are saturated for  the settings $\Sigma:
|\psi\rangle = |0\rangle; \hat{a}=\sigma_z, \hat{a}^\prime = \sigma_x;
\hat{b}  =\sigma_z$.   It may  be  noted  that  there is  no  backward
signaling  from Bob to  Alice, i.e.,  Alice's outcome  probabilties as
derived from  $P(x,y|\hat{a},\hat{b})$ above are  independent of Bob's
settings, as expected.

Now   $P(x,y|\hat{a},\hat{b})$    yields   the   correlator   $\langle
\hat{a}\hat{b}  \rangle  =  \sum_{x,y} xyP(x,y)  =  \frac{1}{2}\langle
\{\hat{a}\hat{b}\}  \rangle =  \vec{a}\cdot\vec{b}$, where  $\hat{a} =
\vec{a}\cdot\vec{\sigma}$  and  $\hat{b} =  \vec{b}\cdot\vec{\sigma}$.
This correlator (for qubits)  is independent of temporal order, though
there is a  signaling from Alice to Bob.   These temporal correlations
are the same as that obtained by von Neumann measurements on singlets.
Both CHSH  and LG inequalities are microscopically  violated, with the
Cirelson   bound   \cite{cir80}  being   the   same   in  both   cases
\cite{fritz10}.

If the  state is the maximally  mixed $I/2$, then $s=\tau  = 0$, since
$\hat{a}^2  =   \left(\hat{a}^\prime\right)^2  =  I$   for  any  qubit
observable   with  spectrum  $\pm   1$.  Interestingly,   because  the
correlators  $\langle \hat{a}\hat{b}  \rangle$  are state-independent,
this state will nevertheless violate the LG inequality maximally.

\begin{centering}
\begin{figure}
\includegraphics[width=7cm]{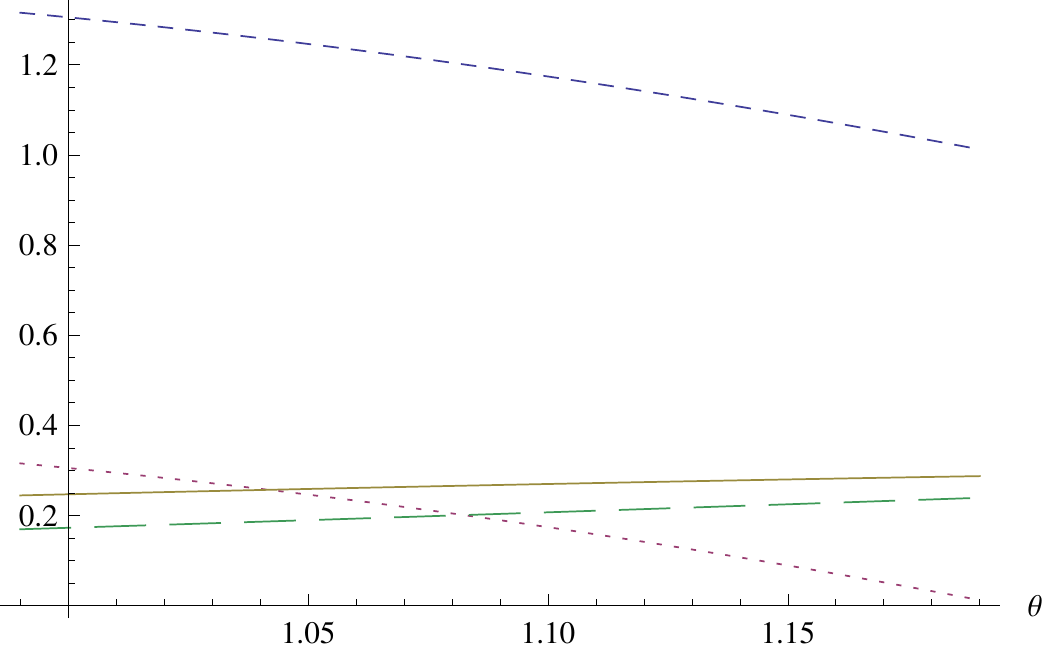}
\caption{For the range of angles  $\theta$ given, the LG inequality is
  always   violated,   with   normalized   LG   inequality   violation
  ($\Lambda(\textbf{P})/\Lambda_0$, small-dashed line) exceeding 1. In
  the region  where signaling ($S(\textbf{P})$  with maximization over
  $b$  restricted to $\{\hat{b},\hat{b}^\prime\}$,  large-dashed line)
  exceeds  the  disturbance  cost  ($C_\Lambda$, dotted  line),  above
  $\theta  \gtrsim  1.08$, the  LG  inequality  violations  in QM  are
  classical, according to Eq.  (\ref{eq:geneq}).  The plain line marks
  the Holevo bound (asymptotically accessible signal).}
\label{fig:bell}
\end{figure}
\end{centering}

\paragraph{Non-classicality.} 
We  propose  that  a  bipartite correlation  \textbf{P}  is  classical
precisely  if 
\begin{equation}
S(\textbf{P}) =  C(\textbf{P}),
\label{eq:cla}
\end{equation}
for  in this  case the  available signal  can be  used in  a classical
mechanism  to simulate  the  correlation, and  conversely a  classical
scheme for  simulation exists  only if this  equality holds  (cf.  Eq.
(\ref{eq:metamath})).   The $\textbf{d}^{j_0}$  strategies,  for which
$S(\textbf{P}) =  C(\textbf{P}) = 0$, are  trivially classical.  Thus,
correlation   \textbf{P}   is   nonclassical   if   $S(\textbf{P})   <
C(\textbf{P})$.  In conjunction with Eq.  (\ref{eq:piro}), a necessary
and sufficient condition for classicality is
\begin{equation}
S(\textbf{P}) \ge C_\Lambda(\textbf{P})
\label{eq:geneq}
\end{equation}
applicable generally to spatial, temporal and contextual correlations.
In the  spatial (i.e., Bell) case,  by no-signaling $S(\textbf{P})=0$,
so that Eq.   (\ref{eq:geneq}) is equivalent to $C_\Lambda(\textbf{P})
\le 0$, which is just  the usual CHSH inequality.  Thus, any violation
of the  CHSH inequality is necessarily  non-classical. More generally,
the form (\ref{eq:geneq}) coincides  with the usual forms of Bell-type
and contextuality  inequalities, but  deviates for LG  inequalities on
account  of   the  non-commutativity  of  the   correlated  terms.   A
non-trivial  instance of this  can be  seen in  Figure \ref{fig:bell},
where  $\hat{a},  \hat{b},  \hat{a}^\prime$ and  $\hat{b}^\prime$  are
oriented  in  the  $xz$  plane,  and separated  by  angular  intervals
$\theta$; the  initial state  is the +1  eigenstate of  $\hat{a}$. The
range $\theta > 1.08$  is classical by criterion (\ref{eq:geneq}) even
though the LG inequality is violated.

A    PR     box    maximally    violates     (\ref{eq:geneq})    while
$\textbf{d}^{\prime0_1}  \equiv \delta_a^x\delta_a^y$, which  is 1-bit
signaling  but does  not violate  inequality  (\ref{eq:lg}), maximally
satisfies  it.  We  define the  \textit{signal deficit},  $\eta \equiv
C(\textbf{P})-   S(\textbf{P})$,    which   vanishes   for   classical
correlations.  If inequality  (\ref{eq:geneq}) is violated, then $\eta
=   C_\Lambda(\textbf{P})   -   S(\textbf{P})$   by  virtue   of   Eq.
(\ref{eq:piro}),  and  represents  the   number  of  bits  on  average
necessary over  the available  signaling to simulate  the disturbance.
For  a  given correlation  inequality,  it  quantifies  the degree  of
nonclassicality.

For  sequential  measurements  on  a  qubit,  $S(\textbf{P})$  in  Eq.
(\ref{eq:chiX})  is  bounded above  by  the  Holevo  quantity $\chi  =
S(\alpha\rho_0 +  \beta\rho_1) - \alpha S(\rho_0)  - \beta S(\rho_1)$.
Direct  substitution in  Eq. (\ref{eq:chiX})  using  settings $\Sigma$
yields          $S(\textbf{P})          =         H(\alpha)          +
\alpha\log\left(\frac{2\alpha}{1+\alpha}\right)                       +
\frac{1-\alpha}{2}\log\left(\frac{1-\alpha}{1+\alpha}\right), $ which,
when maximized, yields $S(\textbf{P}) = \mu_s \equiv \log(5)-2 \approx
0.32$ bits  at $\alpha=3/5$.  This is the  maximum signaling quantumly
possible from  Alice to Bob  \cite{fritz10}, and saturates  the Holevo
bound.   Since  $C_\Lambda  \approx  0.41$ bits  for  maximal  quantum
violation  of   inequality  (\ref{eq:lg})   and  hence  also   the  LG
inequality, the violation  of the inequality (\ref{eq:geneq}) follows.
Thus maximal \textit{quantum} violation of inequality (\ref{eq:lg}) is
indeed  non-classical.   It  entails  a  signal  deficit  of  $\eta  =
0.41-0.32=0.09$ bits.

Setting $S(\textbf{P}) = \mu_s$ in Eq. (\ref{eq:geneq}), we find that
\begin{equation}
\Lambda(\textbf{P})  \le   2(\mu_s  +   1)  \approx  2.64,
\end{equation}   
as the  signal-corrected version  of the LG  inequality (\ref{eq:lg}),
whose  violation  gives a  sufficient  condition for  nonclassicality.
That   its  violation   is  not   necessary  is   clear   from  Figure
\ref{fig:bell}, in the approximate range $\theta \in [1.0,1.03]$.

By  Eq.   (\ref{eq:geneq}),   the  maximally  nonlocal  and  signaling
correlations $\textbf{d}^{0_1}$  and $\textbf{d}^{3_1}$ are classical,
but cannot be  accessed deterministically because intrinsic randomness
ensures that complements  in a signaling pair occur  always in tandem.
Since  this coupling  behavior is  the cause  of  non-vanishing signal
deficit,  intrinsic  randomness  emerges  as the  ultimate  origin  of
nonclassicality.

For the  maximal quantum  violation of the  CHSH and  LG inequalities,
$\eta \approx  0.41$ and $0.09$,  indicating that the  maximal quantum
violation of the CHSH inequality  is in a sense more nonclassical than
the maximal quantum violation of the LG inequality.  Interestingly, as
we saw earlier, for the maximally mixed state $S(\textbf{P})=0$.  Thus
the maximal violation  of the LG inequality with  an initial maximally
mixed state is  more non-classical than that by  an initial pure state
according to this criterion.

Intuitively, a qubit is  a non-classical object.  However, Bell showed
that the outcome of a single projective measurement on any qubit state
can be classically simulated  with a HV model \cite{bell66}.  Further,
a  proof  of  non-classicality  via contextuality  requires  dimension
greater than  2, which  thus does  not apply to  a qubit  subjected to
projective  measurements. However,  our  discussion above  establishes
non-classicality of a qubit by demonstrating temporal correlations for
which $\eta > 0$.

\textit{Intrinsic randomness, no-cloning, etc.}   It can be shown that
nonclassical  features   like  randomness,  no-cloning,   privacy  and
monogamy of  correlations, and  the impossibility of  perfect cloning,
etc.,  which occur  in non-signaling  nonlocal  theories \cite{mag06},
also occur  in signaling correlations that  are nonclassical according
to  the criterion  (\ref{eq:geneq}),  though their  strength tends  to
diminish in the measure  that signaling in the correlations increases.
These details  will be presented  elsewhere, and here we  indicate the
general sense in  which signaling undermines nonclassicality.  Suppose
Alice  and Bob  share an  `unbalanaced  PR box',  i.e., a  correlation
$\textbf{Q}(p)   =  p\textbf{d}^{0_1}   +   (1-p)\textbf{d}^{3_1}$  by
combination of two  signal complements.  Defining intrinsic randomness
by $I[\textbf{Q}(p)]  \equiv \min\{p,1-p\}$, which  we take to  be $p$
without loss  of generality. Thus,  we find $s  + 2I = 1$,  implying a
trade-off between  signaling and randomness.  More generally, $s  + 2I
\ge 1$  if we include  $d^{\prime j_1}$ strategies; cf.   Appendix and
also Ref.  \cite{hall10}).  When  a system is nonclassical, then $s<1$
and intrinsic randomness is necessarily non-vanishing.

Irrespective of $p$, $\textbf{Q}(p)$ satisfies $a \cdot \overline{b} =
x  \oplus  y$.    With  perfect  cloning  by  Bob,   one  has  $a\cdot
\overline{b}^\prime  =   x  \oplus   y^\prime$,  so  that:   $a  \cdot
(\overline{b} \oplus  \overline{b}^\prime) = y  \oplus y^\prime$, from
which  Bob determininstically  obtains  Alice's input.   On the  other
hand,  $S(\textbf{Q}(p))  =  1-2p$.   Generalizing no-cloning  to  the
stipulation of  `no-cloning above  available signaling', we  find that
perfect cloning violates  the generalized no-cloning by $1  - (1-2p) =
2p$ bits. Thus  the violation is maximum (and  hence most `forbidden')
when    the     correlations    are    most     non-classical,    viz.
$\textbf{Q}(\frac{1}{2})$, with $\eta=1$; and least when the system is
classical, viz. $\textbf{Q}(0)$ with $\eta=0$.

\paragraph{Nonclassicality and  metamathematical incompleteness.}  
We now mention briefly the similarity  of our main result to a theorem
in metamathematics, the study of mathematics using mathematical tools.
(More  generally,   a  \textit{meta-theory}   is  a  theory   about  a
theory). At the moment, the analogy is admittedly very sketchy, but we
feel  that it  offers  potential clarity  in  the way  we think  about
physical  theories  by taking  on  an  `outside  view'.  A  well-known
metamathematical  result  is that  of  incompleteness  due to  G\"odel
\cite{kgodel},   according   to   which   given   any   axiomatization
$\textbf{A}$  of   arithmetic,  if  it  is  consistent,   then  it  is
\textit{incomplete},  in the  sense of  there being  truths (theorems)
expressible in $\textbf{A}$ but  not provable within $\textbf{A}$.  An
existential  proof  of the  result  is that  the  set  of theorems  in
$\textbf{A}$  has  the  cardinality  of  the continuum,  i.e.,  it  is
uncountably  large, whilst  the  number of  proofs  is only  countably
infinite.

Strictly speaking, while signaling $S(\textbf{P}$) is a concept in the
base   theory    (QM),   communication   cost    $C(\textbf{P})$)   is
meta-theoretic,  as   therefore  are  Eqs.    (\ref{eq:metamath})  and
(\ref{eq:geneq}),  which are  statements \textit{about}  QM  `from the
outside'  rather than  from \textit{within}  QM.  Thus  the  notion of
non-classicality  as  developed here  is  also meta-theoretic.   Given
theory  $\mathbb{T}$, predictions  in it  are,  technically, theorems,
while signaling indicates  a sequence of causes and  effects, which is
like  a  train  of logical  inferences,  and  thus  is like  a  proof.
Accordingly:   \textit{completeness}   (resp.,   \textit{consistency})
entails  that all  (resp. only)  predicted effects  in the  theory are
explainable  via   the  available  signaling.    Thus  nonclassicality
corresponds  to  incompleteness  in   that  there  exist  effects  not
attributable to signals. A little  thought shows that it is related to
EPR  incompleteness \cite{epr35}.  The  analogy of  nonclassicality to
metamathematical incompleteness  gives the sense  that nonclassicality
is generic rather than pathological.

\paragraph{Discussion and conclusions.} 

Our  work  provides  a  new  characterization  of  nonclassicality  of
bipartite correlations, and thus of  a theory that allows it, in terms
of the signaling available within.  We attribute non-classicality to a
single  unified   postulate:  that  of  signal   deficit  (instead  of
nonlocality and  no-signaling), which  in turn arises  ultimately from
intrinsic randomness.  Our work furnishes a framework to go beyond the
no-signaling  paradigm, leading to  new questions,  such as  ``Why are
signaling  correlations   always  local  in  \textit{non-relativistic}
QM?'', and  to make a  clear separation between theory  and metatheory
within physics, highlighting the  possible primacy of information as a
fundamental resource of Nature.

\paragraph{Acknowledgments.}
SA acknowledges  support through the INSPIRE  fellowship [IF120025] by
the Department of Science and Technology, Govt. of India.

\bibliography{quantum}

~\\

\hrule

\appendix

\section{Appendix of supplementary material}

\hrule
~\\

\textbf{Derivation of Eq. (2).} \\

We have:
\begin{eqnarray}
I(A:Y) &=& H(\alpha) - H(A|Y) \nonumber \\
 &=& H(\alpha) - \sum_{a=0,1}\sum_{y = \pm}P(a,y)\log(P(a|y)),
\label{eq:eq}
\end{eqnarray}
where $P(a,y)  = P(y|a)P(a)$ and  $P(a|y) = P(y|a)P(a)/P(y)$.   We let
$P^b_0  \equiv  P(y=+|a=0),  P^b_1  \equiv  P(y=+|a=1),  Q^b_0  \equiv
P(y=-|a=0),    Q^b_1    \equiv    P(y=-|a=1)$   and    $P(a=0)=\alpha,
P(a=1)=\beta$. Further,  $\overline{P} \equiv P(y=+) =  \alpha P^b_0 +
\beta P^b_1$  and $\overline{Q} \equiv  P(y=-) = \alpha Q^b_0  + \beta
Q^b_1$.    Substituting  these  values   in  Eq.   (\ref{eq:eq}),  and
maximizing over $\alpha$ and settings $b$, we obtain Eq. (2).

~\\
\hrule
~\\

\textbf{Communication cost of signaling correlations.} \\

\textbf{Claim.}  \textit{For  any bipartite two-settings, two-outcomes
  correlation}  \textbf{P},  $C(\textbf{P}) =  \max\left(S(\textbf{P}),
  C_\Lambda(\textbf{P})\right)$.

\textbf{Proof.}   Ref.  \cite{pir03}  proves  that $C(\textbf{P})  \ge
C_\Lambda(\textbf{P})$,   and   furthermore   that  $C(\textbf{P})   =
C_\Lambda(\textbf{P})$  if $S(\textbf{P})  = 0$.   Here  we generalize
this    result    by    first    showing   that    $C(\textbf{P})    =
C_\Lambda(\textbf{P})$  if $S(\textbf{P})  \le C_\Lambda(\textbf{P})$.
Consider  the  most general  strategy  with  which  \textbf{P} can  be
constructed  by   using  the  eight   deterministic  local  strategies
$d^{j_0}$  and  the  eight  deterministic 1-bit  strategies  $d^{j_1}$
strategy, given in Eqs.   \ref{eq:d0} and \ref{eq:d1}, in the notation
of Ref. \cite{pir03}.  The local strategies are:
\begin{eqnarray}
\textbf{d}^{0_0} &=& \delta^x_0\delta^y_0;~~ \textbf{d}^{1_0} =  \delta^x_a\delta^y_0 \nonumber \\
\textbf{d}^{2_0} &=& \delta^x_0\delta^y_{\overline{b}};~~ \textbf{d}^{3_0} =  \delta^x_{\overline{a}}\delta^y_{\overline{b}} \nonumber \\
\textbf{d}^{4_0} &=& \delta^x_a\delta^y_b;~~ \textbf{d}^{5_0} =  \delta^x_1\delta^y_b \nonumber \\
\textbf{d}^{6_0} &=& \delta^x_{\overline{a}}\delta^y_1;~~ \textbf{d}^{7_0} =  \delta^x_1\delta^y_1 \nonumber \\
\label{eq:d0}
\end{eqnarray}
and the 1-bit strategies that violated inequality (\ref{eq:lg}) by 4 are:
\begin{subequations}
\begin{eqnarray}
\textbf{d} ^{0_1}  &=&   \delta^x_0\delta^y_{a\cdot\overline{b}};~~~ \textbf{d}^{3_1}  =
\delta^x_1\delta^y_{a\cdot\overline{b} +  1}; \label{eq:03} \\\textbf{d}^{1_1}
&=&   \delta^x_{\overline{a}}\delta^y_{a\cdot  b   +  1}; \textbf{d}^{2_1}  =
\delta^x_{a}\delta^y_{a\cdot   b};   \label{eq:12}   \\\textbf{d}^{4_1}   &=&
\delta^x_{a \cdot  \overline{b}}\delta^y_{0}; ~~~\textbf{d}^{7_1} = \delta^x_{a
  \cdot  \overline{b} +1}\delta^y_{1};  \label{eq:47}  \\\textbf{d}^{5_1}  &=&
\delta^x_{\overline{a} \cdot \overline{b}}\delta^y_{\overline{b}}; ~~~
\textbf{d}^{6_1}            =            \delta^x_{\overline{a}           \cdot
  \overline{b}+1}\delta^y_{b},\label{eq:56}
\end{eqnarray}
\label{eq:d1}
\end{subequations}
Given $\textbf{P} \equiv P(x,y|a,b)$, we wish to construct a protocol
\begin{eqnarray}\label{eq:prob} 
P(x,y|a,b) &=& \sum_{\lambda_0=0}^7q_{\lambda_0}d^{\lambda_0}_{xy|ab}+\sum_{\lambda_1 
=0}^7q_{\lambda_1}d^{\lambda_1 }_{xy|ab} \nonumber \\
  &\equiv& q_0 + q_1.
 \end{eqnarray}
where $q_{j_k}$ are the probabilities for the strategy $d^{j_k}$.  The
probabilities  have to  satisfy normalization  constraints $\sum_{x,y}
P(x,y|a,b)=1$. 

The no-signalling conditions are given as
\begin{eqnarray}\label{eq:nosign} 
\sum_{y}P(x,y|a,b)=\sum_{y}P(x,y|a,b^\prime) \quad \forall b, b^\prime
\nonumber \\ \sum_{x}P(x,y|a,b)=\sum_{x}P(x,y|a^\prime, b) \quad \forall
a,a^\prime
\end{eqnarray}
For the  two-settings two-output  case, taking into  consideration the
normalization  conditions, there are  only 4  independent no-signaling
conditions.  Allowing for their  general violation, these 4 conditions
are:
\begin{widetext}
\begin{subequations}
\begin{eqnarray}
P(-1,y|0,0)   +  P(+1,y|0,0)   &=&  P(-1,y|1,0)   +   P(+1,y|1,0)  \pm
\delta_{03} \label{eq:sig03} \\ P(-1,y|0,1) + P(+1,y|0,1) &=& P(-1,y|1,1)
+  P(+1,y|1,1)   \pm  \delta_{12}  \label{eq:sig12}   \\  P(x,-1|1,0)  +
P(x,+1|1,0)       &=&      P(x,-1|1,1)       +       P(x,+1|1,1)      \pm
\delta_{47} \label{eq:sig47} \\ P(x,-1|0,0) + P(x,+1|0,0) &=& P(x,-1|0,1)
+ P(x,+1|0,1) \pm \delta_{56} \label{eq:sig56},
\end{eqnarray}
\label{eq:sigmn}
\end{subequations}
\end{widetext}
where the $\delta_{j}$'s ($j \in T = \{03, 12, 47, 56\}$) are positive
quantities that  quantify violation of the  no-signaling condition and
satisfy $\sum_{j \in T}\delta_j \le 1$.

Eqs.  (\ref{eq:sig03})  and (\ref{eq:sig12}) represent  signaling from
Alice to  Bob while  Eqs.  (\ref{eq:sig47}) and  (\ref{eq:sig56}) that
from Bob to Alice.  Each such violation of an independent no-signaling
condition  corresponds  to a  pair  of  $d^{j_1}$  strategies of  Eqs.
\ref{eq:d1}, which we call a \textit{signaling pair}.  The two members
of  a given  pair are  signal complements  of each  other.   E.g., the
violation  of no-signaling  in  Eq.  (\ref{eq:sig03})  is produced  by
having  the  complements  in  Eq.  (\ref{eq:03})  occur  with  unequal
probabilities.   More  generally,  the  violation  of  any  of  the  4
independent no-signaling conditions arises when the complements in any
of the signal pairs are imbalanced. Thus:
\begin{eqnarray}
\delta_{03} \equiv |q_{0_1} - q_{3_1}|; && \delta_{12} \equiv |q_{1_1}
- q_{2_1}|,\nonumber\\  \delta_{47}  \equiv  |q_{4_1} -  q_{7_1}|;  &&
\delta_{56} \equiv |q_{5_1} - q_{6_1}|.
\label{eq:deltaj}
\end{eqnarray}

If we  use only the local  and 1-bit strategies  of Eqs. (\ref{eq:d0})
and (\ref{eq:d1}), then
\begin{equation}
S(\textbf{P})  = \max_{j \in T} \delta_j,
\label{eq:delta}
\end{equation}
and
\begin{equation}
C_\Lambda(\textbf{P}) = \sum_{\lambda=0}^7q_{\lambda_1},
\label{eq:sign}
\end{equation}
It   follows  from   Eqs.   (\ref{eq:delta}),   (\ref{eq:deltaj})  and
(\ref{eq:sign}) that  
\begin{equation}
S(\textbf{P}) \le C_\Lambda(\textbf{P}).
\label{eq:bu}
\end{equation}  
Thus we must have
\begin{equation}
\forall_j \delta_j \le C_\Lambda(\textbf{P}).
\label{eq:deltaX}
\end{equation}
We  will return  to the  more  general case  than Eq.  (\ref{eq:sign})
later.

To show that $C(\textbf{P}) = C_\Lambda(\textbf{P})$ for any signaling
\textbf{P} such that $S(\textbf{P}) \le C_\Lambda$, we will provide an
explicit  protocol that  realizes \textbf{P}  using  the deterministic
strategies  $\textbf{d}^{j_0}$  and  $\textbf{d}^{j_1}$ a  correlation
\textbf{P}  which  violates  precisely  one of  the  four  independent
no-signaling conditions  (\ref{eq:nosign}), say Eq.  (\ref{eq:sig03}).
The idea straightforwardly generalizes to the general case of all four
no-signaling   conditions   being   violated.   In   particular,   let
$\delta_{03} > 0$ but the other $\delta_{j}$'s vanish.

The contribution of  the negative sign for $\Lambda  (\textbf {P})$ is
only  from  the $\textbf{d}^{j_0}$  strategies,  and  fixes the  eight
$q_{0_j}$'s.    For  example,   $q_{1_0}  =   P(1,0|1,1)$,  and   so  on
\cite{pir03}.    The  positive   terms  are   constructed   with  both
$\textbf{d}^{j_0}  $ and $\textbf{d}^{j_1}$  deterministic strategies.
For   example,   using    Eqs.    (\ref{eq:d0})   and   (\ref{eq:d1}),
 $P(+1,+1|0,0)  =  q_{0_0}  + q_{1_0}  +  q_{4_0}  +
q_{0_1} + q_{2_1} + q_{4_1} + q_{6_1}$ or $q_{0_1} + q_{2_1} + q_{4_1}
+  q_{6_1} =  P(+1,+1|0,0)  - P(+1,+1|1,0)  - P(-1,+1|1,1)  -
P(+1,-1|0,1)$.   Solving for the $q_{j_1}$'s using the normalization
conditions, the three independent no-signaling conditions and the lone
independent signaling condition, we find
\begin{eqnarray}
q_{0_1}  &=& \frac{1}{2}\left(\frac{C_\Lambda +3\sigma_{03}}{4}\right)
\pm     \frac{\delta_{03}}{2},     \nonumber     \\    q_{3_1}     &=&
\frac{1}{2}\left(\frac{C_\Lambda  +3\sigma_{03}}{4}\right)  \mp  \frac
     {\delta_{03}}{2}       \nonumber       \\       q_{1_1}       &=&
     q_{2_1}=q_{4_1}=q_{5_1}=q_{6_1}=q_{7_1}                          =
     \frac{1}{2}\left(\frac{C_\Lambda          -\sigma_{03}}{4}\right),
     \nonumber \\
\label{eq:signasign}
\end{eqnarray}
where  $0  \le  \sigma_{03}  \le  C_\Lambda$  in  order  to  guarantee
positivity of $q_{j_1}$ ($j=1,2,4,5,6,7$),  and $0 \le \delta_{03} \le
\frac{C_\Lambda}{4}+3\sigma_{03}$ to guarantee positivity of $q_{0_1}$
and  $q_{3_1}$.  The  result of  Ref.   \cite{pir03} for  the case  of
no-signaling  correlations is obtained  as a  special case  by setting
$\delta_{03} = \sigma_{03}=0$.

More generally, any of the  other three no-signaling conditions can be
violated, and we  can define $\sigma_{j}$ ($j \in  \{12, 47, 56\}$) as
above (to determine the total probability of a signal complement), and
the  above  result  can  be  generalized  to  an  arbitrary  signaling
distribution  for which  Eq. (\ref{eq:bu})  holds.   The communication
cost associated with  this protocol with signalling is  given, in view
of Eqs.  (\ref{eq:sign}) and (\ref{eq:signasign}), by $C(\textbf{P}) =
C_\Lambda(\textbf{P})$.

Since  for the  strategies $\textbf{d}^{j_0}$  and $\textbf{d}^{j_1}$,
Eq. (\ref{eq:bu}) holds, it therefore follows that if $S(\textbf{P}) >
C_\Lambda(\textbf{P})$,   then  deterministic   strategies   that  are
signaling  but  produce  no  violation  of  inequality  (1),  such  as
$\textbf{d}^{\prime 0_1} \equiv  \delta^x_a\delta^y_a$, should be used
with    total    probability    $q^\prime_1    =    S(\textbf{P})    -
C_\Lambda(\textbf{P})$.  In simulating the protocol, Bob receives with
probability $S(\textbf{P})  = q_1 +  q^\prime_1$ a 1-bit  message.  To
simulate \textbf{P}, he implements the above protocol with probability
$q_1/S(\textbf{P})$, and  chooses a 1-bit  Bell non-violating strategy
$\textbf{d}^{\prime j_1}$ with probability $q_1^\prime/S(\textbf{P})$,
to ensure  that there is no  violation beyond $C_\Lambda(\textbf{P})$.
The  general  super-disturbance  signaling  correlation  can  thus  be
simulated with $S(\textbf{P})$ bits on average, so that $C(\textbf{P})
= S(\textbf{P})$.

Combining   the   results    for   the   above   sub-disturbance   and
super-disturbance cases of the  signal, we obtain the required result.
\hfill $\blacksquare$

\end{document}